\documentclass[conference]{IEEEtran}
\IEEEoverridecommandlockouts

\usepackage{cite}
\usepackage{amsthm,amsmath,amssymb}
\usepackage{tikz}
\usepackage{pgfplots}
\usetikzlibrary{shapes,arrows}
\usetikzlibrary{tikzmark}

\usepackage{algorithm}
\usepackage{algorithmicx}
\usepackage{algpseudocode}
\usepackage{cite}
\usepackage{enumitem}
\usepackage{hyperref}
\newtheorem*{remark}{{Remark}}
\hypersetup{
    colorlinks=true,
    linkcolor=blue,
    filecolor=magenta,      
    urlcolor=blue,
    pdftitle={Quantum ACPF},
    citecolor = blue,
    }

\usetikzlibrary{arrows.meta}
\usetikzlibrary{positioning,fit,backgrounds}
\usetikzlibrary{decorations.pathmorphing,patterns}

\usepackage{tikz}
\usepackage{tikz-network}

\begin{document}

\title{Conditions for Quantum Advantage in AC Power Flow}

\author{Parikshit Pareek$^{1}$$^\dagger$, Abhijith Jayakumar$^{2}$, Carleton Coffrin$^{2}$, Sidhant Misra$^{2}$\\
$^{1}$Department of Electrical Engineering, Indian Institute of Technology Roorkee, India\\
$^{2}$Los Alamos National Laboratory, New Mexico, USA

\thanks{\textsuperscript{$\dagger$}Corresponding Author \texttt{pareek@ee.iitr.ac.in}. Pareek acknowledges the funding support provided by the ANRF PM Early Career Research Grant (ANRF/ECRG/2024/001962/ENS) and the IIT Roorkee Faculty Initiation Grant (IITR/SRIC/1431/FIG-101078)}
}


\maketitle

\begin{abstract}
This paper aims to contextualize the requirements for Quantum Computing (QC) algorithms to achieve a quantum advantage in solving the alternating current power flow (ACPF) problem, with a focus on runtime complexity. First, we establish a benchmark for a QC iterative solver to demonstrate an advantage over the classical Newton-Raphson Load Flow (NRLF) algorithm. Next, we derive a baseline expression for the end-to-end runtime complexity of any Gate-based QC algorithm as $\Omega(N \kappa/\varepsilon),$ reflecting dependence on system size $N$, condition number $\kappa$, and error tolerance $\varepsilon$. Finally, we highlight key areas where QC algorithms may offer potential benefits over NRLF in addressing the standard ACPF problem.
\end{abstract}





\section{Introduction}

\IEEEPARstart{S}{olving} alternating current power flow (ACPF) is one of the most fundamental operations of electrical energy system analysis. The set of nonlinear ACPF equations are solved for system state, nodal voltage and line flow values, for given network structure and loading conditions \cite{john1994power}. Two major computational challenges in solving the ACPF problem, comes due to nonlinearity of the ACPF equations and size of the power grids. Additionally, with increase in renewable and distributed energy resources, solving ACPF under uncertainty for probabilistic assessment poses additional computational challenges \cite{prusty2017critical,hasan2019existing}.
Recently, quantum power flow (QPF) algorithms have focused on linear formulations (direct current power flow (DCPF) and fast-decoupled load flow (FDLF)), claiming to leverage QC's asymptotic speedup potential over classical methods \cite{qpf,9122420,saevarsson2022quantum,liu2022quantum,golestan2023quantum}. These QPF studies rely on QC's potential to provide exponential speedup for solving linear equations via Harrow, Hassidim, and Lloyd (HHL) algorithm \cite{harrow2009quantum}. However, recent work \cite{pareek2024demystifying} demonstrates through an end-to-end complexity analysis that HHL-based QPF methods are slower than classical algorithms (conjugate gradient (CG)), for solving linear power flow problems.


In this paper, we address two key questions: \textit{What will it take for a quantum ACPF algorithm to outperform NRLF?} and \textit{What is the best possible runtime complexity achievable by a quantum ACPF algorithm?} To answer the first, we analyze NRLF's runtime complexity under convergent cases in Section \ref{sec:nrlf}. For the second, we derive a lower bound on the complexity of quantum ACPF algorithms in Section \ref{sec:baseline}. 
Finally, we highlight a few potential challenging directions where quantum ACPF methods could provide advantages over traditional approaches like NRLF.




\section{NRLF Complexity Benchmark}\label{sec:nrlf}
Formulating equations for nodes with known active or reactive power injections reduces the ACPF problem to solving a nonlinear system of equations \cite{john1994power}. Thus, iterative methods are widely used for ACPF, with NRLF being the most adopted \cite{milano2010power}. The NRLF works on the principle of first order expansion, and at each iteration, the updates in node voltage values (both magnitude and angle) are calculated by solving a linear system of equations \cite{john1994power}, for which Conjugate Gradient (CG) is the most efficient algorithm by exploiting sparsity and $\varepsilon$-convergence instead of complete solve \cite{shewchuk1994introduction}. 
However, unlike in the case of DCPF, the Jacobian in ACPF is not positive definite \cite{john1994power}. Thus CG must be run  on the normal equations

\begin{align}
A\mathbf{x} = \mathbf{b} \implies A^TA \mathbf{x} = A^T\mathbf{b} \implies M\mathbf{x} = \mathbf{c}
\end{align}

\begin{figure}[t]
    \centering
    \includegraphics[width=\columnwidth]{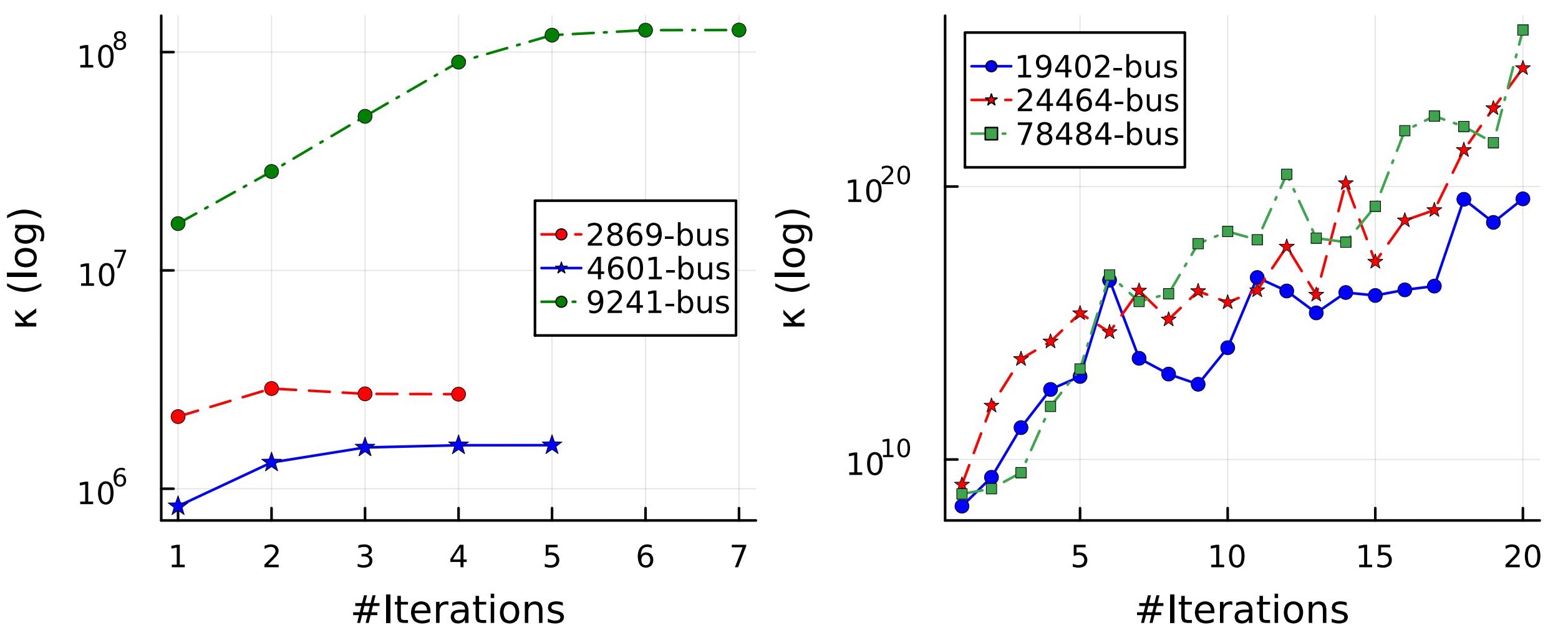}
    \caption{Condition number ($\kappa$) variations for the \texttt{PGLib} \cite{pglib} power transmission network datasets reveal potential numerical challenges in solving ACPF. \textbf{Left:} Variation in the Jacobian condition number in \textit{converging instances} for medium-scale systems. It is clear that the condition number increases with each iteration before converging to the final value. \textbf{Right:} Variation in the Jacobian condition number across 20 NRLF iterations for large-size systems in \textit{non-converging instances}. It shows that the Jacobian condition number varies significantly between iterations and increases rapidly for non-converging instances, thus necessitating use of $\kappa = \max_i \kappa_i$ in \eqref{eq:nrlf}. Note that, condition number of initial Jacobian grows as polynomial, with the power system size ($N$ Bus), due to its direct dependence on admittance bus matrix which has similar relationship \cite{pareek2024demystifying}. }
    \label{fig:condition}
\end{figure}

Important observations related to this conversion is that $M = A^TA$ is symmetric for any matrix $A$ and condition number of  $M$ is the square of the condition number of $A$ i.e. $\kappa(M) = \kappa^2(A)$. Therefore, the runtime complexity of NRLF, using CG for normal equations based linear system solve is\footnote{For a symmetric positive definite matrix $A$, the CG runtime complexity is $\mathcal{O}(Ns\sqrt{\kappa}\log(1/\varepsilon_c))$ \cite{shewchuk1994introduction}. Also, note that Jacobian size in NRLF is of order $2N$ but constant does not alter runtime complexity expression. For more on Big-O notation, readers can refer: \href{https://web.mit.edu/16.070/www/lecture/big_o.pdf}{Link}.}
\begin{align}\label{eq:nrlf}
   \mathcal{O}(HNs\kappa\log(1/\varepsilon_c)) 
\end{align}
Here, $H$ is the maximum number of iterations taken by NRLF to solve ACPF, and $N$ is the number of buses in the system. Further, $s$ is the maximum number of non-zero values in a row i.e. sparsity which remains constant across iterations (changes due to bus-type stitching are marginal thus can be ignored from runtime complexity arguments). The \(\varepsilon_c\) denotes the `smallest' energy norm error \cite{pareek2024demystifying,shewchuk1994introduction} achieved in any iteration of the linear solve process of NRLF using the CG method. Furthermore, the condition number \(\|A^{-1}\|\|A\|\) of the Jacobian varies throughout the iterations of NRLF, as shown in Fig. \ref{fig:condition}. Since runtime complexity represents the worst-case scenario, we define the overall condition number as \(\kappa = \max_i \kappa_i\) where $i=1\dots H$ and \(\kappa_i\) is the condition number at the \(i\)-th iteration of the Jacobian. 


The $Ns$ term in the CG complexity is the complexity of performing a sparse matrix-vector multiplication. Here, $A^TA$ may not be a sparse matrix even if $A$ is sparse. However, the dependence on sparsity remains unchanged due to the fact that we never calculate $A^TA$ explicitly and conjugate descent is calculated as $A^T[A\mathbf{d}]$ (with $\mathbf{d}$ being steepest descent direction) which results in two $\mathcal{O}(Ns)$ matrix-vector multiplication steps. Due to definition and structure of Jacobian non-zero elements in any row cannot be greater than two times of those in admittance bus ($\rm Y$-Bus) \cite{john1994power,milano2010power} and $\rm Y$-Bus sparsity doesn't scale with system size and shown to remains below 30 for majority of the \texttt{PGLib} systems \cite{pareek2024demystifying,pglib}. Therefore, sparsity can be considered in the form of a constant, and removed from the runtime complexity, reducing \eqref{eq:nrlf} as $\mathcal{O}(HN\kappa\log(1/\varepsilon_c)) $.

It is imperative to understand that NRLF does not provide any upper bound on $H$ and only guarantees the convergence if the initial guess is `close' to the final solution \cite{milano2010power}. Also notice that, $\varepsilon_c$ is the error in linear system solving (in this case $M\mathbf{x} = \mathbf{c}$) not in the NRLF convergence. The  final error in the NRLF solution is implicitly fixed by $H$. We determine this $H$ empirically in \figureautorefname~\ref{fig:condition} only for systems where we can guess an appropriate initial solution such that NRLF converges.   
Moreover, in practice, FDLF is considered to be the faster alternative to the NRLF, which takes more iterations but less time per iteration. The reduction in iteration time is due to reduced Jacobian size, obtained via decoupling \cite{milano2010power}. The runtime complexity of FDLF is $\mathcal{O}(K_fNs\sqrt{k}\log(1/\varepsilon_c))$ (or $\mathcal{O}(K_fN\sqrt{k}\log(1/\varepsilon_c))$ with sparsity as non-scaling variable) where $K_f$ is the maximum iterations required in FDLF\footnote{The FDLF doesn't work with Jacobian's but uses static susceptance matrices. Thus, we do not need to convert these to normal equations \cite{pareek2024demystifying}.}. 
Further, absence of convergence guarantee for arbitrary starting point with NRLF means, in industry practice, ACPF solvers use variants of NRLF with divergence control and \textit{robustification heuristics} \cite{powerworldsimulator,milano2010power}.  Lastly, we have empirical evidence that number of NRLF iterations does not scale with system parameters such as size (see left plot of Fig. \ref{fig:condition} and NRLF related discussion in \cite{john1994power}). 
Thus, considering the number of iterations $H$'s invariability with system size, we can obtain a reduced form of the complexity expression \eqref{eq:nrlf} in terms of system size, condition number, and error as $\mathcal{O}(N\kappa\log(1/\varepsilon_c))$. 

In \eqref{eq:nrlf}, error term $\varepsilon_c = \sqrt{\mathbf{e}^TM\mathbf{e}}$ is the energy norm error (from CG convergence), with $\mathbf{e}$ being absolute difference between current and final solution \cite{shewchuk1994introduction}. On the other hand, error in quantum computing is measured using $\ell_2$-norm between the normalized solutions i.e. $\varepsilon = \sqrt{\mathbf{e}^T\mathbf{e}}/\| x\| $ \cite{harrow2009quantum}. Therefore, below we convert classical complexity in terms of $\ell_2$-norm, to establish direct comparable upper bound.

As we solve the normal equation, the resultant matrix $M = A^TA$ is always positive definite. We adopt the normalization convention by which $\|b \| = 1$ This changes the solution by a constant factor and can be undone after $x$ is obtained. Under these conditions, it can be shown that $  \frac{\varepsilon}{ \kappa(A)} \leq \varepsilon_c$. 
From these considerations, the per iteration complexity of NRLF will be
\begin{align}\label{eq:nrlf_ub}
    \mathcal{O}(N\kappa \log(\kappa/\varepsilon))
\end{align}
Thus, \eqref{eq:nrlf_ub} is the complexity barrier which a quantum ACPF algorithm must cross in order to show any signs of potential quantum advantage In the following section, we present an optimistic Q-ACPF complexity constructed as a lower-bound of any end-to-end algorithm pipeline for solving ACPF using quantum linear solvers.

\begin{figure*}[t]
    \centering
    \begin{align*}
\textbf{C:}  \quad   
& \mathcal{O}(\textcolor{blue}{H} \textcolor{red}{Ns} \textcolor{red}{\kappa} \textcolor{red}{\log(\varepsilon_c^{-1})}) 
\xrightarrow[\text{independent}]{\text{H is N}}  
\mathcal{O}(\textcolor{red}{Ns} \textcolor{red}{\kappa} \textcolor{red}{\log(\varepsilon_c^{-1})}) 
\xrightarrow[N \, \text{independent}]{\text{Sparsity is}}  
\mathcal{O}(\textcolor{red}{N} \textcolor{red}{\kappa} \textcolor{red}{\log(\varepsilon_c^{-1})}) 
\xrightarrow{\log(\kappa/\varepsilon) \geq \log(\varepsilon_c^{-1})}  
\mathcal{O}(\textcolor{red}{N} \textcolor{red}{\kappa} \textcolor{red}{\log(\kappa/\varepsilon)}) \\ \quad \\
\textbf{Q:} \quad  
& \mathcal{O}(\textcolor{blue}{HT_r} (\textcolor{blue}{T_p} + \textcolor{blue}{T_s})) 
\xrightarrow[H=1]{\text{QRAM}}  
\Omega(\textcolor{blue}{T_r} (\textcolor{red}{\log(N)} + \textcolor{blue}{T_s})) 
\xrightarrow[\text{Lower Bound}]{\text{State Propagation}}  
\Omega(\textcolor{blue}{T_r} (\textcolor{red}{\log(N)} + \textcolor{red}{\kappa})) 
\xrightarrow[\text{Tight Bound}]{\text{Read Out}}  
\Omega(\textcolor{red}{N} \textcolor{red}{\kappa} \textcolor{red}{\varepsilon^{-1}})
    \end{align*}
    \caption{\textbf{Top:} NRLF complexity bound considering independence of sparsity and iterations $H$ with respect to system size $N$ and error bounds. \textbf{Bottom:} Lower bound of quantum complexity considering QRAM availability, state propagation, readout and optimistic consideration of single iteration solve.}
    \label{fig:complexity}
\end{figure*}

\begin{remark}
Recently, the Holomorphic Embedding Load Flow Method (HELM) was proposed to solve ACPF independently of initial values \cite{7923960}, but it is slower than NRLF for achieving the same precision \cite{7923960}. Note that, a hybrid method combining HELM for initialization and NRLF for convergence could be explored, where \eqref{eq:nrlf} will represents the lower bound, considering HELM is slower than NRLF \cite{7923960}. Therefore, we argue that for a Quantum ACPF algorithm, NRLF's complexity is the benchmark to surpass for demonstrating quantum advantage.
\end{remark}


\section{Optimistic Quantum-ACPF Complexity}\label{sec:baseline}
A quantum ACPF solver will have three sequential components, contributing to runtime: i) state preparation (reading the problem), ii) state propagation (quantum algorithm) and iii) tomography (reading the solution) (See Fig. 2 of \cite{pareek2024demystifying} for more details). Following this, lets develop a lower bound on runtime complexity of any Quantum-ACPF solver.  

Let us consider a standard ACPF problem where the input is the specified input vector \(\mathbf{x} \in \mathbb{R}^{2N}\), and the output is the vector \(\mathbf{y} \in \mathbb{R}^{2N}\) (including voltage magnitudes and angles).  To assess end-to-end complexity, we first examine \textbf{state preparation complexity} $T_p$. Quantum linear system solvers like HHL require state preparation via \textit{amplitude encoding }\cite{schuld2018supervised}. Since an arbitrary load vector lacks a well-defined generative function, an \textit{amplitude embedding} circuit is needed to prepare the injection vector state\footnote{NRLF solves linear systems using specified power values: \(\texttt{specified} - \texttt{calculated} = \texttt{Jacobian} \times \texttt{voltage update}\) \cite{john1994power,milano2010power}.}. The runtime complexity for this operation is \(T_p = \Theta(N)\), considering complexity for preparing a general quantum state \cite{dalzell2023quantum}. If we assume access to  quantum random access memory (QRAM), state preparation scales as $\log N$ \cite{dalzell2023quantum}. Hereafter, we optimistically assume QRAM availability and consider $T_p = \Theta(\log N)$.

Now we consider \textbf{state propagation (quantum algorithm) complexity} $T_s$.  It was first shown by authors in \cite{harrow2009quantum} that a quantum linear system solver has a query complexity lower bound of $\Omega(k)$, i.e. any quantum linear solver  must take time that at least scales linearly with the condition number. \emph{Query complexity} here refers to the number of calls the algorithm makes to a quantum circuit that holds the data for $A$. Hence the actual runtime scaling can be worse if a large quantum circuit is needed to encode $A$.  In the spirit of finding a optimistic lower bound, assume the ideal scenario for the quantum solution time $T_s = \Theta(\kappa)$. We will also assume that the number of iterations required to solve the full non-linear system doesn't scale with system parameters, as we did with the classical solver.  
 
Lastly, we consider \textbf{readout complexity} $T_r$ which will be the last stage of any proposed Quantum ACPF algorithm, convert the final ACPF solution from quantum state to classical state \cite{pareek2024demystifying,dalzell2023quantum}. For ACPF problem, the solution vector is dense containing voltage magnitudes and angles. Since we have access to the state preparation unitaries that prepared the solution vector (basically the circuit we used to solve the problem) we can use the tomography algorithm defined in \cite{van2023quantum} to perform tomography in complexity $\Theta(N/\epsilon)$. Here the $\epsilon$ error is guaranteed in the $\ell_2$ norm, and defines the error between true classical output and output read via tomography. 
 
Now, to arrive at complete end-to-end complexity we need to combine these three components of complexity in the manner in which QC algorithms work. Due to the fundamental axioms of quantum mechanics, a single readout operation generates only one sample from the quantum state and destroys it in the process\cite{pareek2024demystifying,dalzell2023quantum,van2023quantum}. Therefore, the whole process of state preparation and state propagation need to be done $T_r$ times. This implies that generalized end-to-end complexity expression will be $\Theta(T_r(T_p+T_s))$ or $\Omega(N\kappa/\varepsilon)$, where $\Omega(\cdot)$ reflects that it is a lower bound complexity.

Note that we have assumed that the solution of one quantum linear system solve can be fed into the next without any extra overhead. In practice, we might require extra tomography after each linear system solve to set up the system for the next solve.

\subsubsection*{\textbf{Quantum Advantage Regime}}

Now using the lower bound we have for the quantum solver and the upper bound for the classical solver, we look for a regime where quantum advantage can exist

Analyzing final complexity expressions in Fig. \ref{fig:complexity} reveals that the factors $N$ does not impact the quantum-classical complexity ratio. Thus, the scaling with respect to the error tolerance is the primary determinant of potential quantum advantage, i.e., $\log(\kappa/\varepsilon)$ for classical versus $1/\varepsilon$ for the quantum algorithm. To illustrate, consider a stringent error $\varepsilon = 10^{-6}$ and the high $\kappa = 10^8$, that we observed in \figureautorefname \ \ref{fig:condition}. Then the ratio of quantum to classical complexity evaluates to
$\frac{1/\varepsilon}{\log(\kappa/\varepsilon)} \approx \frac{1}{10^{-6} \cdot \log 10^{14}} \approx 3.1 \times 10^4$, indicating that the quantum cost is substantially higher than the classical cost. For a moderate error requirement of $\varepsilon = 10^{-3}$, the ratio reduces to $Q/C \approx 39.$ For this value of $\kappa$, this ratio  goes below unity only around $\varepsilon \approx 0.05$, which is a fairly large relative error in the solution.

This $Q/C$ ratio that we are computing here computed using assumptions that are heavily in favor of the quantum algorithm and the high $\kappa$ we chose for the illustration. This indicates that in most practical situations, the $1/\varepsilon$ dependence of the quantum algorithm precludes the possibility of any meaningful quantum advantage. This highlights two narrow possibilities for potential quantum advantage. First, in scenarios with exceedingly coarse accuracy demands, the quantum algorithm could approach parity or slightly outperform classical methods. Second, if overhead constants of classical NRLF are much higher compared to the quantum algorithm to compensate for the effect of logarithmic error scaling.


    



\section{Potential Quantum ACPF Frontiers and Conclusions}  
This paper presents a comparative analysis of classical NRLF runtime complexity and lower bounds on quantum complexity. The results indicate that quantum computing for ACPF  can only  be justified in situations where higher error rates are acceptable or where classical overheads could offset the logarithmic error-scaling advantage of classical methods. Beyond direct speedup over NRLF, quantum approaches might offer promising avenues for addressing long-standing challenges in ACPF: enumerating multiple solutions to capture diverse operational states \cite{mehta2016numerical}, handling ill-conditioning and detecting bifurcation points as in Continuation Power Flow methods, and mitigating initialization sensitivity to improve convergence robustness. Any potential advantage however, requires these methods to achieve lower runtime complexity than classical alternatives, such as Holomorphic-NRLF, performing the same tasks. These considerations delineate the frontiers where quantum ACPF algorithms may meaningfully complement or surpass classical approaches.

\bibliographystyle{IEEEtran}
\bibliography{main}

\appendix 
\section{Error transforms}
Notice that quantum and classical linear system solvers solve different systems that are related to the original $Ax = b$ system. Moreover, their error guarantees differ. 

\paragraph{Classical system}: The system solved by CG is $M \mathbf{x} = \mathbf{c}$, where $M = A^TA$ and $\mathbf{c} = A^T\mathbf{b}$. This is because, natively, CG can only solve positive systems. The condition number of $M$ is the square of that of $A$

\paragraph{Quantum system.} Quantum linear solvers natively only handle Hermitian system. Hence we first have to encode the non-symmetric $A$ into a symmetric matrix in the following block diagonal form,
\[
\tilde{A} = 
\begin{pmatrix}
0 & A \\
A^{T} & 0
\end{pmatrix}
\]

Using this we solve the equation,
\begin{equation}
    \tilde{A} \begin{pmatrix}
       0 \\ \mathbf{x}  
    \end{pmatrix} = \begin{pmatrix}
    \mathbf{b} \\ 0
    \end{pmatrix}
\end{equation}
It is easy to see that the eigenvalues of $\tilde{A}$ are the singular values of $A$, and hence their condition numbers match.

\subsection{Error comparison}
Error in the classical algorithm is given by $\varepsilon_c = \sqrt{\mathbf{e}^T M \mathbf{e}}$. The error in the quantum setting is $\varepsilon = \frac{\| \mathbf{e} \|}{\| \mathbf{x} \|}$

\begin{align}
\varepsilon_c &\geq \sqrt{\lambda_{min}(M)} \ \|\mathbf{e}\|, \\
&= \sigma_{min}(A) \ \| \mathbf{e} \|, \\
&=  \sigma_{min}(A) \ \varepsilon \ \| \mathbf{x} \|, \\
&\geq   \sigma_{min}(A) \ \varepsilon \ \sigma_{min}(A^{-1}), \\
= \varepsilon/\kappa
\end{align}

\end{document}